\documentclass[prl,twocolumn,amsmath,amssymb,nofootinbib]{revtex4} 
\usepackage{graphicx} 
\usepackage{amsmath}
\usepackage{amsfonts,amsbsy}
\usepackage{amssymb}

\usepackage{xcolor}
\definecolor{lcolor}{rgb}{0.5,0,0}
\definecolor{citcolor}{rgb}{0,0.3,0.0}
\usepackage[breaklinks,colorlinks,urlcolor=blue,citecolor=citcolor,linkcolor=lcolor]{hyperref}

\def\be{\begin{equation}}
\def\ee{\end{equation}}
\def\bea{\begin{eqnarray}}
\def\eea{\end{eqnarray}}

\newcommand{\xt}{{\mathbf{x}_T}}
\newcommand{\yt}{{\mathbf{y}_T}}
\newcommand{\kt}{{\mathbf{k}_T}}
\newcommand{\as}{\alpha_{\mathrm{s}}}
\newcommand{\qs}{Q_\mathrm{s}}

\newcommand{\lqcd}{\Lambda_{\mathrm{QCD}}}

\begin{document}

\title{
The distribution of linearly polarized gluons and
elliptic azimuthal anisotropy in DIS dijet production at high energy
}

\author{Adrian Dumitru}
\email{Adrian.Dumitru@baruch.cuny.edu}
\affiliation{Department of Natural Sciences, Baruch College, CUNY,
17 Lexington Avenue, New York, NY 10010, USA}
\affiliation{The Graduate School and University Center, The City University of New York, 365 Fifth Avenue, New York, NY 10016, USA}

\author{Tuomas Lappi}
\email{Tuomas.v.v.Lappi@jyu.fi}
\affiliation{
Department of Physics, %
 P.O. Box 35, 40014 University of Jyv\"askyl\"a, Finland
}
\affiliation{
Helsinki Institute of Physics, P.O. Box 64, 00014 University of Helsinki,
Finland
}

\author{Vladimir Skokov}
\email{VSkokov@bnl.gov}
\affiliation{Department of Physics, Western Michigan University, Kalamazoo, MI 49008, USA}
\affiliation{RIKEN/BNL Research Center, Brookhaven National Laboratory, Upton, NY 11973, USA}

\begin{abstract}
We determine the distribution of linearly polarized gluons of a dense
target at small $x$ by solving the B-JIMWLK rapidity evolution
equations. From these solutions we estimate the amplitude of $\sim \cos 2\phi$
azimuthal asymmetries in DIS dijet production at high energies. We
find sizeable long-range in rapidity azimuthal
asymmetries with a magnitude in the range of $v_2=\langle\cos
2\phi\rangle\sim 10\%$.
\end{abstract}

\maketitle

Transverse momentum dependent (TMD) 
factorization~\cite{Collins:1981uw,Angeles-Martinez:2015sea} in
deep inelastic scattering predicts a distribution for linearly
polarized gluons in an unpolarized target~\cite{Mulders:2000sh,Meissner:2007rx}.
This is reflected in $\cos 2\phi$ asymmetries in dijet
production~\cite{Dominguez:2011wm,Metz:2011wb} and in other
processes~\cite{Boer:2009nc,Boer:2010zf,Qiu:2011ai}. 
To date little is known about the
magnitude of these functions in the small-$x$ regime of high
energies. In this paper we perform first estimates of these functions
by solving the B-JIMWLK renormalization group 
equations~\cite{Balitsky:1995ub,Balitsky:1998kc,Balitsky:1998ya,Jalilian-Marian:1997jx,Jalilian-Marian:1997gr,JalilianMarian:1997dw,Kovner:1999bj,Kovner:2000pt,Iancu:2000hn,Iancu:2001ad,Ferreiro:2001qy,Weigert:2000gi}.
Also, we use our solutions to analyze the magnitude of the resulting
$\cos 2\phi$ asymmetry in dijet
production~\cite{Dominguez:2011wm,Dominguez:2011br} at leading order.
These could be tested at a future electron-ion collider
(EIC)~\cite{Boer:2011fh,Accardi:2012qut}, where the small-$x$ effects
discussed here can be enhanced by using a nuclear target.

Recent data for high multiplicity
p+p~\cite{Khachatryan:2010gv,ATLAS_ridge_pp13TeV} and
p+Pb~\cite{CMS:2012qk,Chatrchyan:2013nka,Khachatryan:2015waa,Abelev:2012ola,Abelev:2014mda,Aad:2014lta,Aad:2013fja}
data at the LHC have revealed long-range (in rapidity) angular $\cos
2\phi$ ``ridge'' correlations in particle production high multiplicity
events.  The magnitude of these long range correlations is
conventionally parametrized in terms of $v_2\equiv\langle \cos
2\phi\rangle$. In fact, the azimuthal correlation in DIS dijet
production at high energy originates also from the long-ranged eikonal
interaction and so results in a similar experimental signature as the
``ridge''.  To make this connection explicit we shall parametrize
the azimuthal structure arising from the linearly polarized gluon
distribution in terms of $v_2=\langle\cos 2\phi\rangle$, and determine its
dependence on the rapidity imbalance of the dijet.

\begin{figure*}[htb]
\centerline{
\includegraphics[width=0.45\linewidth]{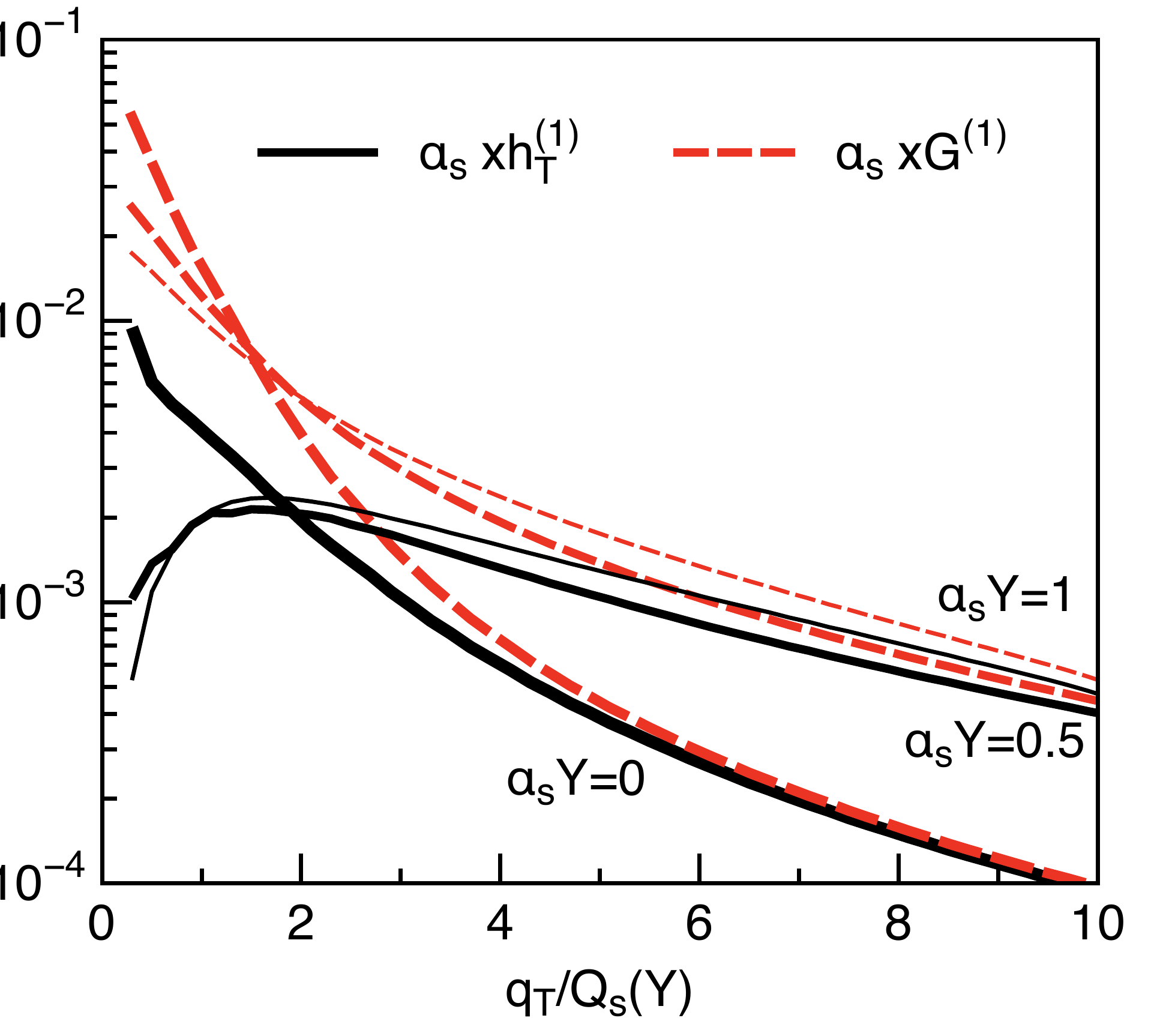}
\includegraphics[width=0.45\linewidth]{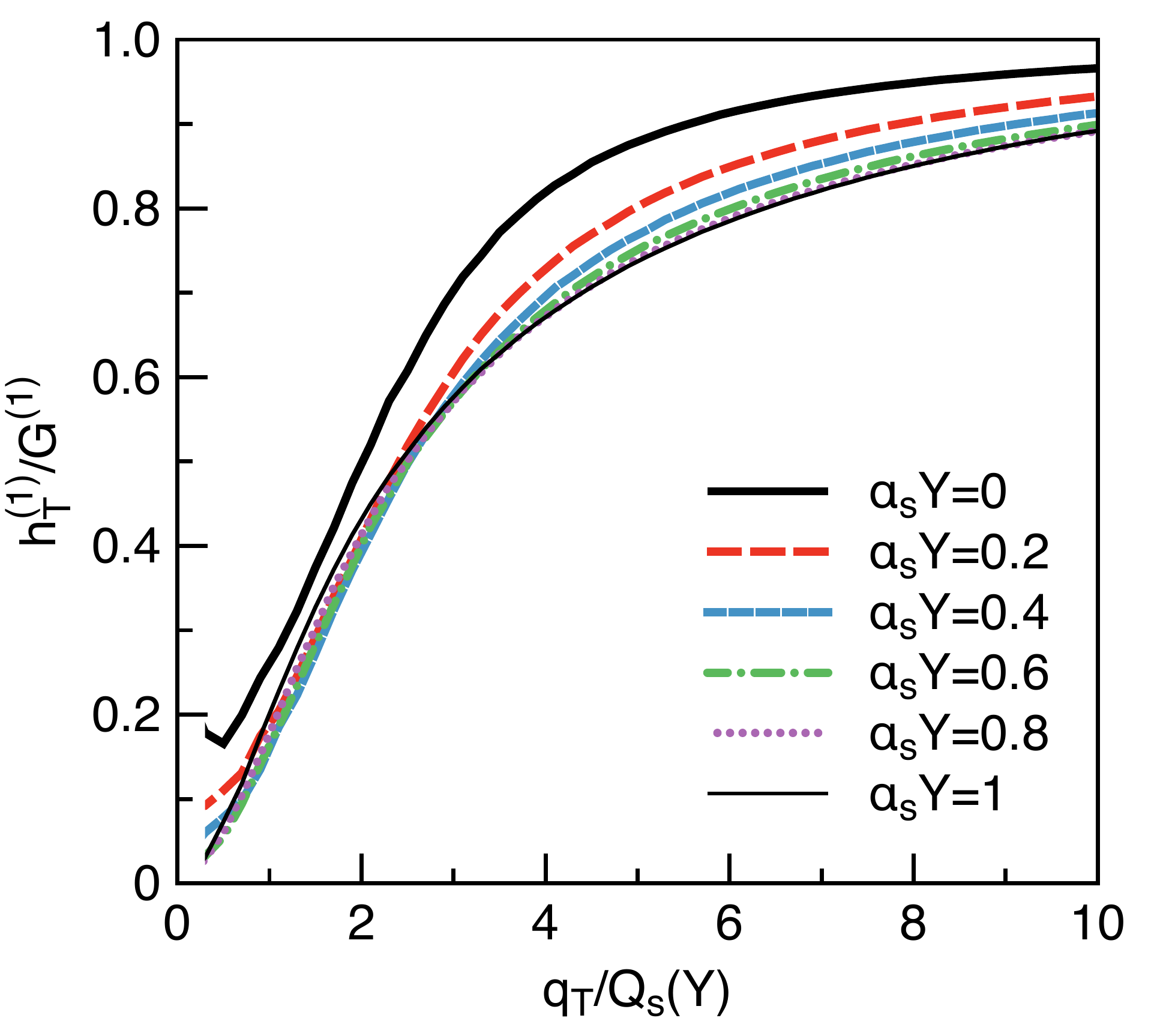}}
\caption{Linearly polarized and unpolarized WW gluon distributions
  versus transverse momentum $q_\perp$ at different rapidities
  $Y$. Transverse momentum is measured in units of the saturation
  momentum $\qs(Y)$. The curves correspond to evolution at fixed $\as
  = 0.15$.
}
\label{fig:xG1_xh1}
\end{figure*}

At leading order the cross section for inclusive production of a dijet in
$\gamma^*$-nucleus scattering is given by~\cite{Metz:2011wb,Dominguez:2011wm}
\begin{widetext}
\begin{eqnarray}
E_1E_2
\frac{d\sigma ^{\gamma _{T}^{\ast }A\rightarrow
    q\bar{q}X}}{d^3k_1d^3k_2 d^2b}
&=&\alpha _{em}e_{q}^{2}\alpha _{s}\delta \left( x_{\gamma ^{\ast
}}-1\right) z(1-z)\left( z^{2}+(1-z)^{2}\right) \frac{\epsilon _{f}^{4}+%
{P}_{\perp }^{4}}{({P}_{\perp }^{2}+\epsilon _{f}^{2})^{4}}
\notag \\ 
&&
\quad \quad \quad \quad \quad \quad 
\times \left[ xG^{(1)}(x,q_{\perp })-\frac{2\epsilon _{f}^{2}{P}%
_{\perp }^{2}}{\epsilon _{f}^{4}+{P}_{\perp }^{4}}\cos
  \left(2\phi\right)xh_{\perp }^{(1)}(x,q_{\perp })\right] ~,
\label{dipoledis2} \\
E_1E_2
\frac{d\sigma ^{\gamma _{L}^{\ast }A\rightarrow q\bar{q}X}}{d^3k_1d^3k_2 d^2b}
&=&\alpha _{em}e_{q}^{2}\alpha _{s}\delta \left( x_{\gamma ^{\ast
}}-1\right) z^{2}(1-z)^{2}\frac{8\epsilon _{f}^{2}{P}_{\perp }^{2}}{(
{P}_{\perp }^{2}+\epsilon _{f}^{2})^{4}}  \notag \\
&&
\quad \quad \quad \quad \quad \quad 
\times \left[ xG^{(1)}(x,q_{\perp })+\cos \left(2
  \phi\right)xh_{\perp }^{(1)}(x,q_{\perp })\right]~. 
\label{dipoledisl2}
\end{eqnarray}
\end{widetext}
Here,
\be
\vec{P}_{\perp } = (1-z)\vec{k}_{1} - z \vec{k}_{2}~~,~~
\vec q_\perp = \vec{k}_1+\vec{k}_2
\ee
are the dijet transverse momentum scale $\vec{P}_{\perp }$ and the
transverse momentum imbalance $\vec{q}_\perp$, respectively. The
transverse momenta of the produced quark and anti-quark are given by
$\vec{k}_1$ and $\vec{k}_2$ and their respective light-cone momentum
fractions are $z$ and $1-z$; the dijet invariant mass is given by
$M=P_\perp/\sqrt{z (1-z)}$. Also, $\epsilon_f^2=z(1-z)Q^2$ with $Q^2$
of order $P_\perp^2$. Here, we restrict ourselves to kinematic configurations
where $\vec{P}_{\perp }$ is greater than $\vec{q}_\perp$, referred to as
the ``correlation limit'' in Refs.~\cite{Dominguez:2011wm,Dominguez:2011br}.

In Eq.~(\ref{dipoledisl2}) $\phi$ denotes the azimuthal angle between
$\vec{P}_{\perp }$ and $\vec{q}_{\perp }$, respectively. We introduce
the following measure for the azimuthal anisotropy,
\be \label{eq:Def_v2}
v_2 \equiv \left< \cos 2 \phi\right>~.
\ee
The average over $\phi$ in this equation is performed with the
weight~(\ref{dipoledis2}) or~(\ref{dipoledisl2}), respectively. Since
\be
x = \frac{1}{s}\left( q_\perp^2 + \frac{1}{z(1-z)}P_\perp^2\right)
\ee
is independent of $\phi$, for a longitudinally/transversally polarized
photon we have
\be
v_2^L = \frac{1}{2} \frac{h_{\perp }^{(1)}(x,q_{\perp
  })}{G^{(1)}(x,q_{\perp })}~~~,~~~ 
v_2^T = - \frac{\epsilon _{f}^{2}{P}_{\perp }^{2}}{\epsilon
  _{f}^{4}+{P}_{\perp }^{4}} \frac{h_{\perp }^{(1)}(x,q_{\perp
  })}{G^{(1)}(x,q_{\perp })}~.
\ee

The linearly polarized $h_{\perp }^{(1)}$ and unpolarized 
$G^{(1)}$ distributions are defined as the traceless part and the trace of the
Weizs\"{a}cker-Williams unintegrated gluon
distribution, respectively:
\begin{equation}
x G^{ij}_{\rm WW}  = \frac{1}{2} \delta^{ij} xG^{(1)}
-\frac{1}{2} \left(\delta^{ij} - 2 \frac{k^i k^j}{k^2} \right) x
h^{(1)}_\perp ~.
\end{equation}
In the CGC framework the gluonic degrees of freedom at small~$x$ are
described by Wilson lines. They are path ordered exponentials in the
strong color field of the target, and cross sections for different
observables can be related to different correlation functions of the
Wilson lines.  The Wilson line is a path ordered exponential of the
covariant gauge field, whose largest component is $A^+$:
\be \label{eq:V_rho} U(\xt) = \mathbb{P} \exp\left\{ ig \int d x^-
A^+(x^-,\xt) \right\}.  
\ee 
The Weizs\"{a}cker-Williams unintegrated gluon
distribution~\cite{Kharzeev:2003wz,Dominguez:2011wm,Dominguez:2011br},
on the other hand, is expressed most naturally
in terms of the light cone gauge
($A^+=0$) field, which has large transverse components. These 
 can be obtained by a gauge transformation 
\be \label{eq:E_WW} {A}^i(\xt) =
\frac{1}{ig} U^\dagger(\xt) \, \partial_i U(\xt) ~.  
\ee
Since, in light cone gauge, the gauge field lives above the light
cone ${A}^i(\xt,x^-)\sim \theta(x^-) {A}^i(\xt)$, this field can also
be thought of as a sheet of color electric field on the light cone
$E^i(\xt,x^-) = \delta(t-z) {A}^i(\xt)$.  The
Weizs\"{a}cker-Williams distribution is simply the two-point
correlator of the light cone gauge fields
\begin{multline}
x G^{ij}_{\rm WW} (x, \vec{k}) =  
\frac{8\pi}{L^2}
\int 
\frac{d^2 \xt} {(2\pi)^2} 
\frac{d^2 \yt} {(2\pi)^2} 
e^{-i \kt \cdot (\xt-\yt)}
\\
\times \left\langle 
A^i_a(\xt)
A^j_a(\yt)
\right\rangle  ,
\label{eq:Def_Gij}
\end{multline}
where we have normalized the distribution with
the transverse area of the target $L^2$. This normalization drops out
of the results expressed in terms of the elliptical asymmetry $v_2$.
For analytical calculations of the functions
$G^{(1)}(x_0,q_{\perp })$ and $h_{\perp }^{(1)}(x_0,q_{\perp })$ in
the McLerran-Venugopalan (MV) model~\cite{McLerran:1994ni,McLerran:1994ka},
see Refs.~\cite{Metz:2011wb,Dominguez:2011br}.

\begin{figure*}[t]
\centerline{
\includegraphics[width=0.45\linewidth]{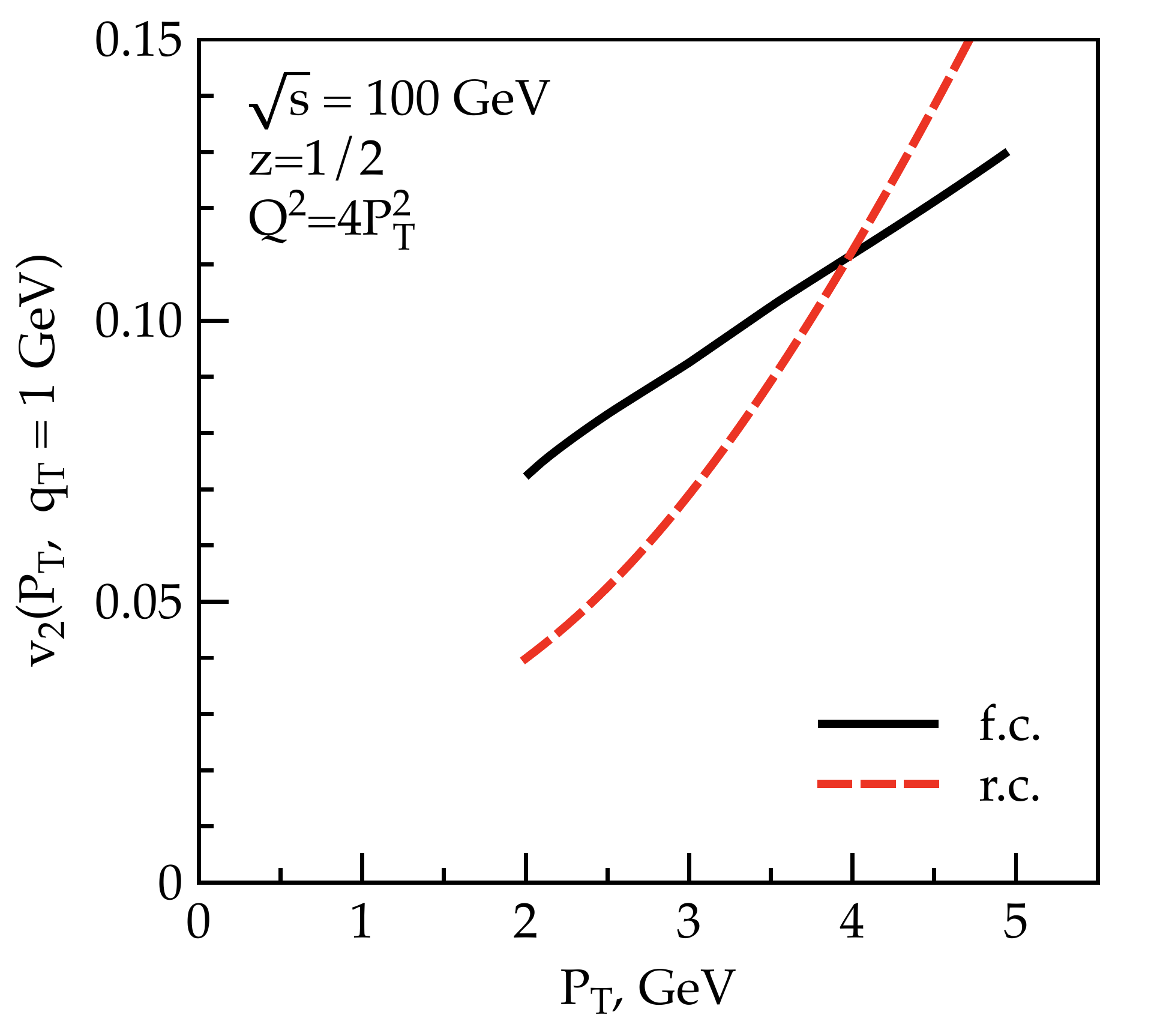}
\includegraphics[width=0.45\linewidth]{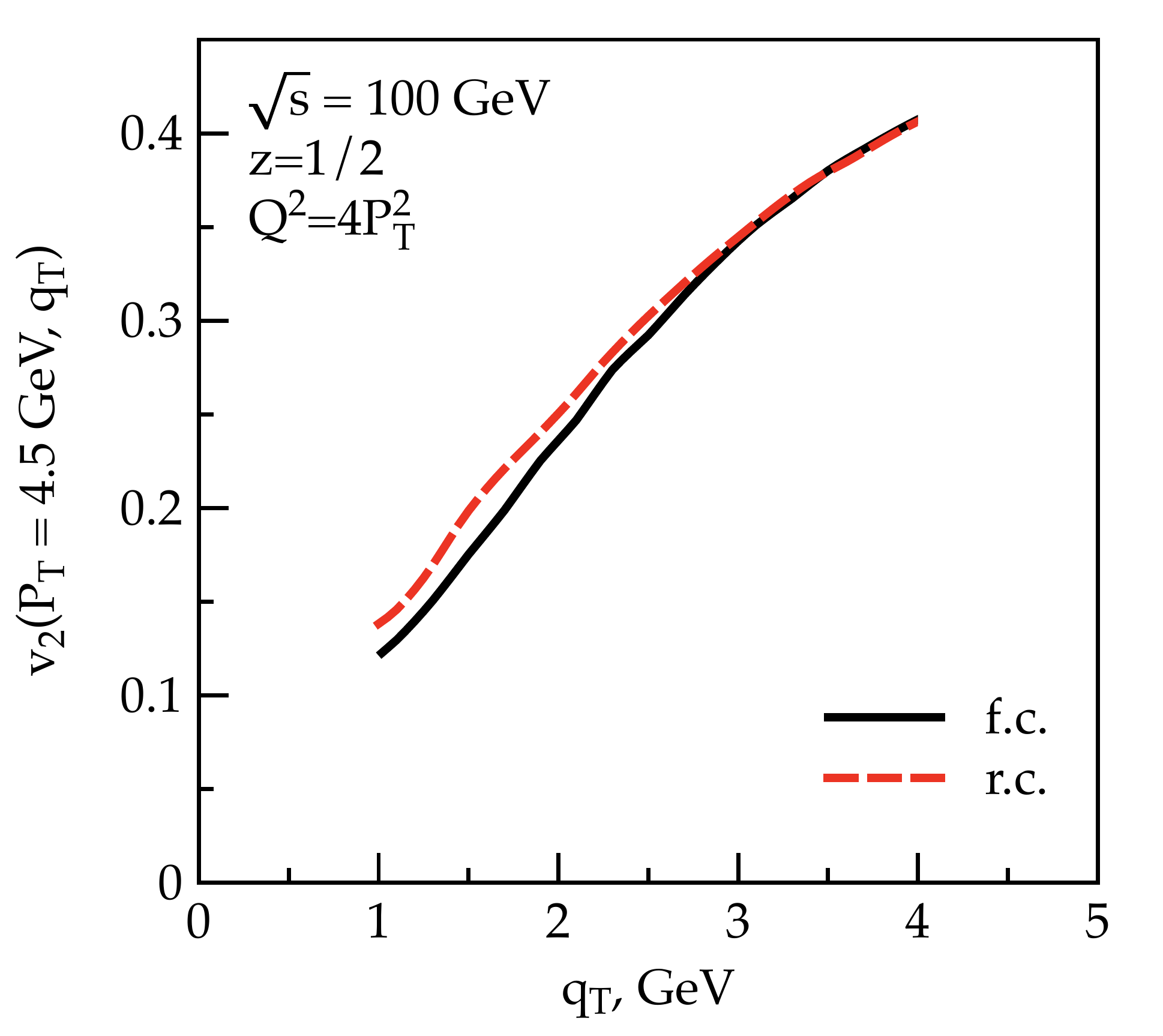}}
\caption{The average azimuthal anisotropy $v_2=\left<\cos 2\phi\right>$
  versus the dijet transverse momentum scale $P_T$ or the dijet
  transverse momentum imbalance $q_T$, respectively. The assumed
  $\gamma^*$A center of mass energy is $\sqrt{s} = 100$~GeV. Since
  $Q^2=4P^2_\perp$ and $z=1/2$ these curves apply to either
  longitudinal or transverse photon polarization. Solid (dashed) lines
  correspond to fixed (running) coupling evolution.
}
\label{fig:v2_qt_Pt}
\end{figure*}

We obtain the Wilson lines $U$ numerically from B-JIMWLK evolution
in $Y=\ln(x_0/x)$,
starting from an initial condition at $x_0=10^{-2}$ using the the MV
model.  The initial condition on the lattice is constructed as
described in detail in Ref.~\cite{Lappi:2007ku}. The B-JIMWLK equation can be
solved on the lattice with a Langevin
method~\cite{Blaizot:2002np,Rummukainen:2003ns}.  We use here the
``left-right'' symmetric~\cite{Kovner:2005jc} numerical method
introduced in Ref.~\cite{Lappi:2012vw}, using either fixed coupling or a
running coupling with the algorithm of Ref.~\cite{Lappi:2012vw}.  As in
e.g. Ref.~\cite{Dumitru:2014nka}, we determine the saturation scale
$\qs$ numerically from the two-point (dipole) function of the Wilson lines.
The renormalization group evolution increases $\qs$ roughly as 
$\qs^2 \sim x^{-0.3}$.
For the calculation of the light cone gauge field one needs Fourier
transforms of derivatives of Wilson lines. Some care must be exercised
to obtain the proper momentum space distribution: we have used two
different centered difference methods (discretizing over one or two
lattice spacings) and found that the results are equivalent.  For the
fixed coupling evolution we take $\as=0.15$ to provide an evolution
speed roughly in line with inclusive HERA data. For running coupling
we use in this preliminary study the slightly overestimated value
$\qs(x_0)/\lqcd = 11$, which also slows down the evolution closer to
experimentally observed values.

For our numerical estimates below we take $Q^2=4P_\perp^2$. Hence, for
$z=1/2$, $v_2^L$ and $v_2^T$ have equal magnitude but
there is a relative phase shift of $\pi/2$. The
physical momentum scale is set by the saturation momentum at $x_0$. To
obtain the numerical values in the plots we take $\qs(x_0)=1$~GeV (for
a $q\bar{q}$ dipole).  The saturation momentum corresponds to the
scale where the forward scattering amplitude is of order 1.

We now turn to describe our results. We first show the solution for the
unintegrated gluon distributions before discussing the azimuthal
asymmetry w.r.t.\ the direction of $\vec{q}_\perp$ of the $\gamma^*$A
cross section.

Figure~\ref{fig:xG1_xh1} shows the dependence of $G^{(1)}$ and
$h_{\perp }^{(1)}$ at different evolution rapidities $Y$ on transverse
momentum. We refrain from showing curves for running coupling
evolution since they look very similar. Either one of the TMDs drops
rapidly as a power of $q_\perp$ at high transverse momentum $q_\perp \gg
\qs$ and so they are best measured at $q_\perp$ of order a few times
the saturation scale. For a heavy-ion target the saturation scale is
boosted (on average over impact parameters) by a factor of $\sim
A^{1/3}$~\cite{Kowalski:2007rw} which facilitates such
measurements in a regime of semi-hard $q_\perp$.

The degree of gluon linear polarization is maximal at high transverse
momentum, $h_{\perp}^{(1)} / G^{(1)} \to 1$; the saturation of the
positivity bound of the cross section has also been observed in
perturbative twist-2 calculations of the small-$x$ field of a fast
quark~\cite{Meissner:2007rx,Metz:2011wb}.  On the other hand
$h_{\perp}^{(1)} / G^{(1)} \ll 1$ at low $q_\perp$ which conforms to
the expected power suppression.  
At fixed $q_\perp/\qs(x)$ the ratio of these functions decreases 
rather slowly with rapidity, 
at least after an initial evolution away from
the MV model towards the B-JIMWLK fixed point. 
This means that, because of the growth of $\qs$, the ratio
$h_{\perp}^{(1)} / G^{(1)}$
at fixed transverse momentum $q_\perp$
decreases with rapidity.
Thus the emission of additional small-$x$ gluons reduces the degree
of polarization. Our results show that this effect can quite well be 
parametrized by geometric scaling as a universal function of
$q_\perp/\qs$. 

\begin{figure}[tb]
\centerline{
\includegraphics[width=0.45\textwidth]{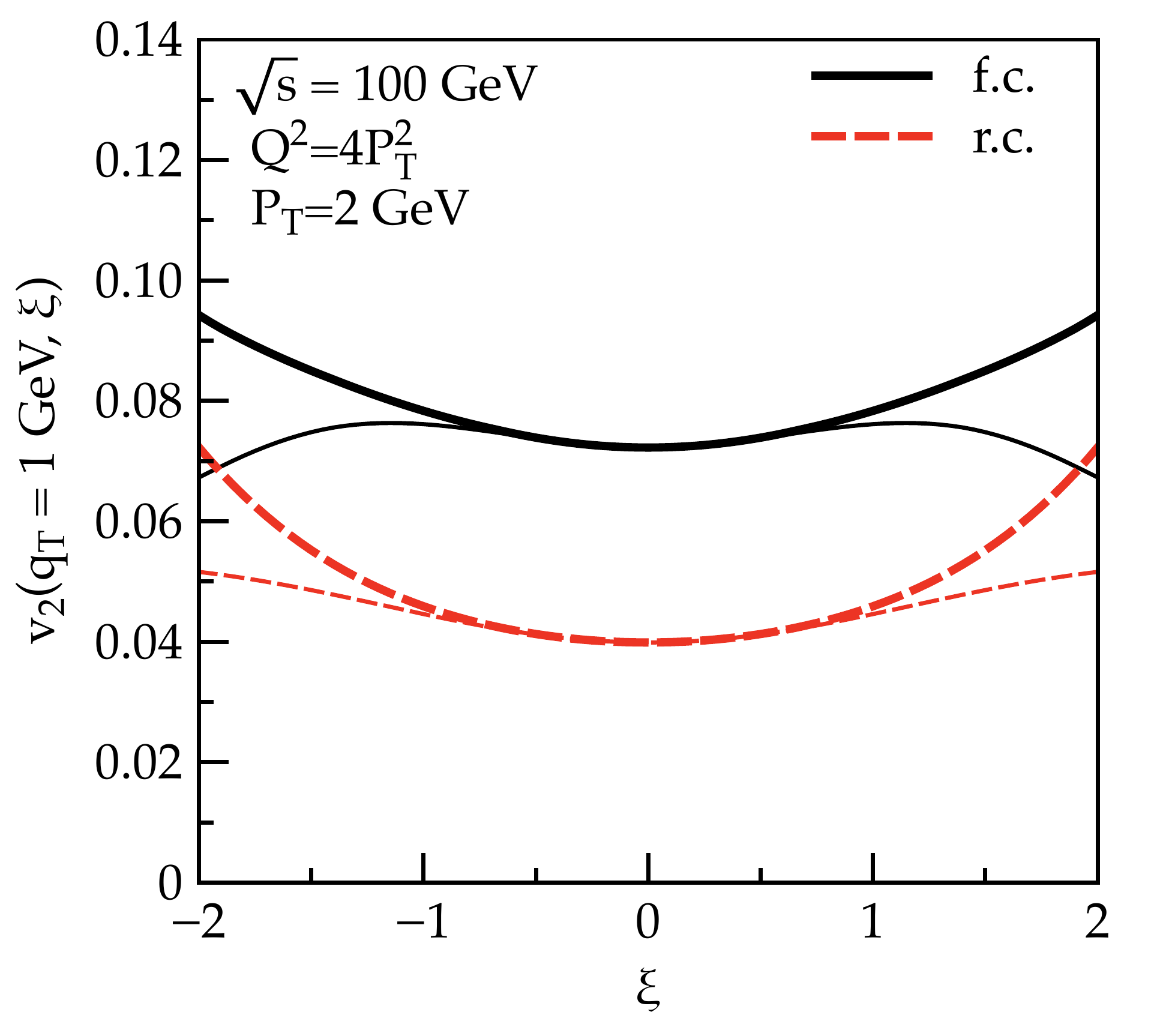}}
\caption{The average azimuthal anisotropy $v_2=\left<\cos 2\phi\right>$
  versus the dijet rapidity imbalance $\xi = \log~(1-z)/z$.
Thick (thin) lines correspond to longitudinal (transverse) photon
polarization.
}
\label{fig:v2_xi}
\end{figure}
In Fig.~\ref{fig:v2_qt_Pt} we show the elliptic asymmetry as a
function of the dijet transverse momentum scale $P_\perp$ and the
transverse momentum asymmetry $q_\perp$. Increasing $P_\perp$
increases $x$ and suppresses evolution effects and so $v_2(P_\perp)$ 
increases towards the MV model initial condition.
The reason for the  difference between the fixed and running coupling curves in 
$v_2(P_\perp)$  is that in this
preliminary study they have not been adjusted to have the same 
evolution speed  $\partial_Y \ln \qs^2(Y)$.
We observe the same
behavior for $v_2(q_\perp)$ even though $x$ increases only slowly with
$q_\perp$; here the increase of the elliptic asymmetry is mainly
due to $h_{\perp}^{(1)}(q_\perp) / G^{(1)}(q_\perp) \to 1$ as
$q_\perp/\qs\gg1$, as shown above. Overall, in the kinematic range
considered in Fig.~\ref{fig:v2_qt_Pt} we find a rather substantial
magnitude of $v_2\sim 10\%$.

Figure~\ref{fig:v2_xi} shows $v_2$ versus the rapidity asymmetry
\be
\xi = \log\, \frac{1-z}{z}~.
\ee
Our calculation applies for moderately large rapidity
separations less than $1/\as$, since we are assuming that 
the two jets are sensitive to the same distribution of Wilson lines.
We find a mild increase of $v_2$ away from $z=1/2$ which is due to the
fact that asymmetric dijet configurations probe the gluon field of the
target at larger values of $x$. The slow evolution of the eikonal
interaction with $x$ translates into a rather flat $v_2(\xi)$ over
several units in $\xi$ away from the boundary of phase space. Hence,
at high energies the azimuthal asymmetry is long range in rapidity.

In summary, we have computed the TMD distribution $h_{\perp}^{(1)}$ of
linearly polarized gluons for a large nucleus at small $x$. We have
used the McLerran-Venugopalan model to obtain initial conditions at
$x_0\sim 10^{-2}$ and the B-JIMWLK equations to evolve to lower $x$.
We find that for realistic values of $x$ and transverse momentum
imbalance $q_\perp$ that $h_{\perp}^{(1)}(x,q_\perp)$ is of
substantial magnitude. This results in large elliptic azimuthal
asymmetries $v_2\equiv\langle\cos 2\phi \rangle \sim 10\%$ in DIS
dijet production. Also, the azimuthal correlations are long range in
rapidity, i.e.\ $v_2$ depends weakly on the rapidity asymmetry $\xi =
\log\, (1-z)/z$.

In the future we intend to check other initial conditions for the evolution,
although we do not expect
qualitative modifications of the results presented here. It will be
interesting, also, to study Sudakov resummation
effects~\cite{Zheng:2014vka,Mueller:2013wwa} as well as more general kinematic
configurations which require quadrupole matrix
elements~\cite{Dominguez:2011wm}.

\begin{acknowledgments}
We are grateful to
E.~Aschenauer, 
A.~Metz, and  
B.~Xiao for useful comments. 
A.D.\ gratefully acknowledges support from the DOE Office of Nuclear
  Physics through Grant No.\ DE-FG02-09ER41620 and from The City
  University of New York through the PSC-CUNY Research Award Program,
  grant 67119-0045.
T.~L.\ is supported by the Academy of Finland, projects 
267321 and 273464. 
V.~S. is supported by RIKEN Foreign Postdoctoral Researcher Program.
This work used computing resources from
CSC -- IT Center for Science in Espoo (Finland) 
and the High Performance Computing Center  at Michigan State University (USA). 

\end{acknowledgments}

\InputIfFileExists{preprint_subm.bbl}

\end{document}